\documentclass[namedreferences]{SolarPhysics}
\usepackage[optionalrh]{spr-sola-addons}
\usepackage{graphicx}
\usepackage{color}                       
\usepackage{url}                         
\def\degree{\ensuremath{^\circ}}

\begin{document}
\begin{article}
\begin{opening}

\title{{\huge \bf  \center DRAFT \\}
	Automated Detection of EUV Polar Coronal Holes During Solar Cycle 23}

\author{M. S.~\surname{Kirk}$^{1}$\sep
        W. D.~\surname{Pesnell}$^{2}$\sep
        C. A.~\surname{Young}$^{1}$\sep
        S. A.~\surname{Hess Webber}$^{3}$
       }

\runningauthor{Kirk et al.}
\runningtitle{Polar Coronal Holes During Solar Cycle 23}

\institute{$^{1}$ ADNET Systems Inc., NASA Goddard Space Flight Center
                     email: \url{mskirk@nmsu.edu} email: \url{c.alex.young@nasa.gov}\\ 
              $^{2}$ NASA Goddard Space Flight Center\\
                     email: \url{william.d.pesnell@nasa.gov} \\
              $^{3}$ Catholic Univ. - Scientific and Engineering Student Intern \\
                     email: \url{hesssh01@gettysburg.edu}        
                     }

\begin{abstract}
A new method for automated detection of polar coronal holes is presented. This method, called perimeter tracing, uses a series of 171, 195, and 304~\AA\ full disk images from the Extreme ultraviolet Imaging Telescope (EIT) on SOHO over solar cycle 23 to measure the perimeter of polar coronal holes as they appear on the limbs. Perimeter tracing minimizes line-of-sight obscurations caused by the emitting plasma of the various wavelengths by taking measurements at the solar limb. Perimeter tracing also allows for the polar rotation period to emerge organically from the data as 33 days. We have called this the Harvey rotation rate and count Harvey rotations starting 4 January 1900.
From the measured perimeter, we are then able to fit a curve to the data and derive an area within the line of best fit. We observe the area of the northern polar hole area in 1996, at the beginning of solar cycle 23, to be about 4.2\% of the total solar surface area and about 3.6\% in 2007. The area of the southern polar hole is observed to be about 4.0\% in 1996 and about 3.4\% in 2007. Thus, both the north and south polar hole areas are no more than 15\% smaller now than they were at the beginning of cycle 23. This compares to the polar magnetic field measured to be about 40\% less now than it was a cycle ago.

\end{abstract}
\end{opening}


\section{Introduction}
	\label{S-Intro} 
The Sun's polar magnetic field has been used to forecast upcoming solar activity for four solar cycles  \cite{Schatten78,Schatten05,Svalgaard01}. The solar magnetic field that extends far beyond the corona into the interplanetary medium is related to the area of the corona threaded by the field. These regions are called solar coronal holes and are defined by both their low emission in EUV and soft x-ray wavelengths and density in the corona as well as their dominantly unipolar magnetic fields \cite{Bohlin77}. First observed in 1941 above the solar limb in visible light \cite{Waldmeier81}, coronal holes are most easily seen in space-based EUV or x-ray images. For ground-based measurements, coronal holes are visible in the equivalent width of the He~{\sc i} 10830~\AA\ line  \cite{Andretta97}. 

Polar coronal holes are coronal holes that occur at the north and south pole of the sun and appear as dark caps on EUV images. Polar holes deserve particular attention because, as the solar cycle progresses toward sunspot number ($R_{\mathrm{Z}}$) maximum, the polar coronal holes become poorly defined and ultimately disappear. During the decline from solar maximum, the polar coronal holes (PCHs) reemerge, becoming better defined and almost symmetric at solar minimum. Due to their relatively long life, PCHs are the source of the majority the solar interplanetary quiet magnetic field. 

\begin{figure}      
   
   \centerline{\hspace*{-0.0\textwidth}
               \includegraphics[width=1.0\textwidth,]{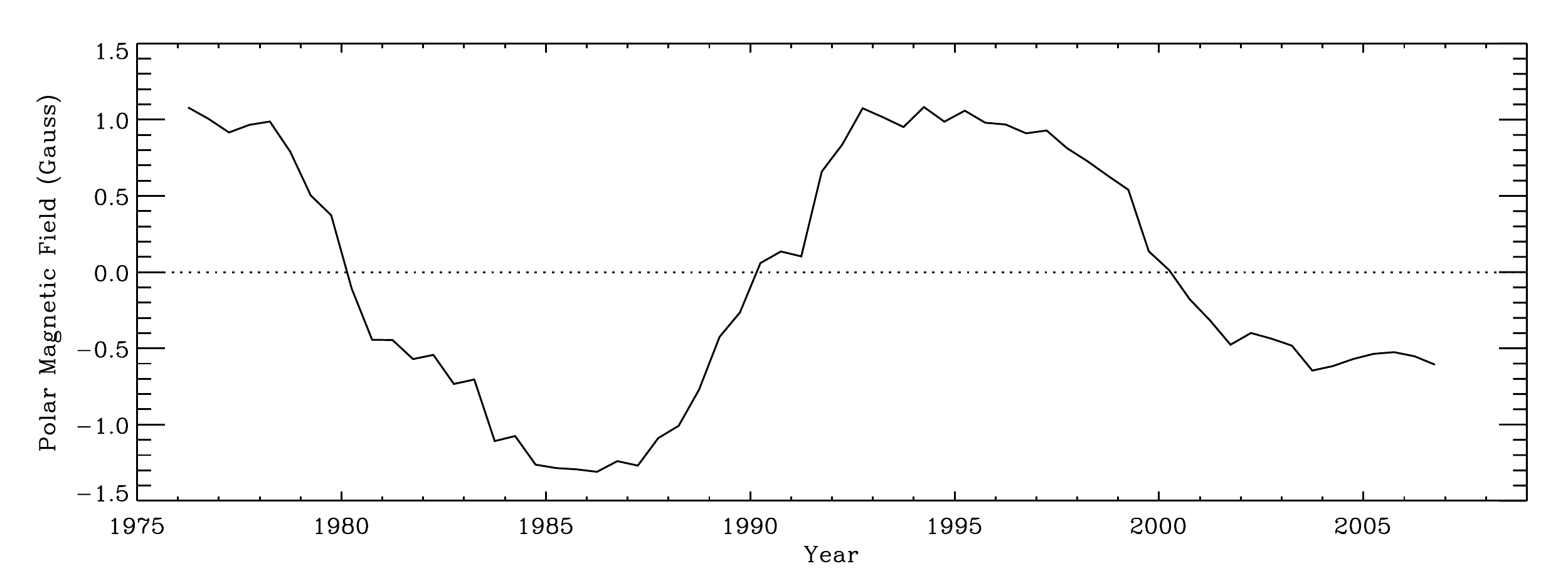}
              }
               \vspace{0.02\textwidth}   
\caption{The average magnetic field measured at WSO beginning in July of 1976 through the end of 2007. Measurements are made every 10 to 14 days of the field at both poles. This plot shows the average of these measurements from both poles, $(B_{\rm N}-B_{\rm S})/2$, smoothed with a 7 point boxcar function to remove the effects of the tilt of the solar rotation axis with respect to the ecliptic. The times when the averaged magnetic field is zero correspond to the polarity flip associated with solar maximum.}
       \label{WSO-figure}
\end{figure}

On average, the polar magnetic field behaves in much the same way as the PCHs. The polar field strength at both poles reaches a maximum at solar minimum. At solar maximum, the polar field decreases through zero on its way to changing its polarity.  Over the past three solar minima the magnitude of the polar field has steadily decreased, indicating a decreasing trend in solar activity. The current solar minimum has a polar field magnitude of 0.6 gauss, compared to 1.0 gauss in the minimum preceding Cycle 23 as shown in Figure~\ref{WSO-figure} \cite{WSO}.

Coronal holes, both polar and equatorial, have historically been identified visually and hand-traced by experienced observers. Examples of these images can be seen in Harvey and Recely (2002) and at the National Solar Observatory (NSO) Synoptic Optical Long-term Investigations of the Sun (SOLIS) project. More recently, there have been a few attempts to automate the process of coronal hole identification and detection \cite{deToma05,Henney05,Malan05,deWit06}. 

Both manual and automatic methods in detecting PCHs have difficulties. Manual detection methods are extremely time consuming and difficult to quantify, as they rely on the observers' qualitative assessments of the images. Many automated methods use synoptic maps for hole detection \cite{Harvey02}. However, synoptic maps do not lend themselves to polar analysis. Synoptic maps are compiled from sections of images of the solar disk, taken over a Carrington rotation (CR) of 27.27 days, and then mapped onto a rectangular grid \cite{Benevol01}. Projecting a sphere into a 2D map is always problematic due to the resulting distortions in the apparent size of surface features; (e.g., the size of Greenland on a Mercator projection of the Earth). On the sun this problem is compounded by solar differential rotation. The polar regions of the sun rotate once every approximately 33 days. Using synoptic maps to investigate polar features forces a CR period onto those features.  Due to these problems, a better method to measure PCHs was developed, as discussed below. 


\section{A New Way of Looking at PCHs}
	\label{S-Justification}
\begin{figure}     
   \centerline{\hspace*{-0.0\textwidth}
               \includegraphics[width=0.9\textwidth,clip=]{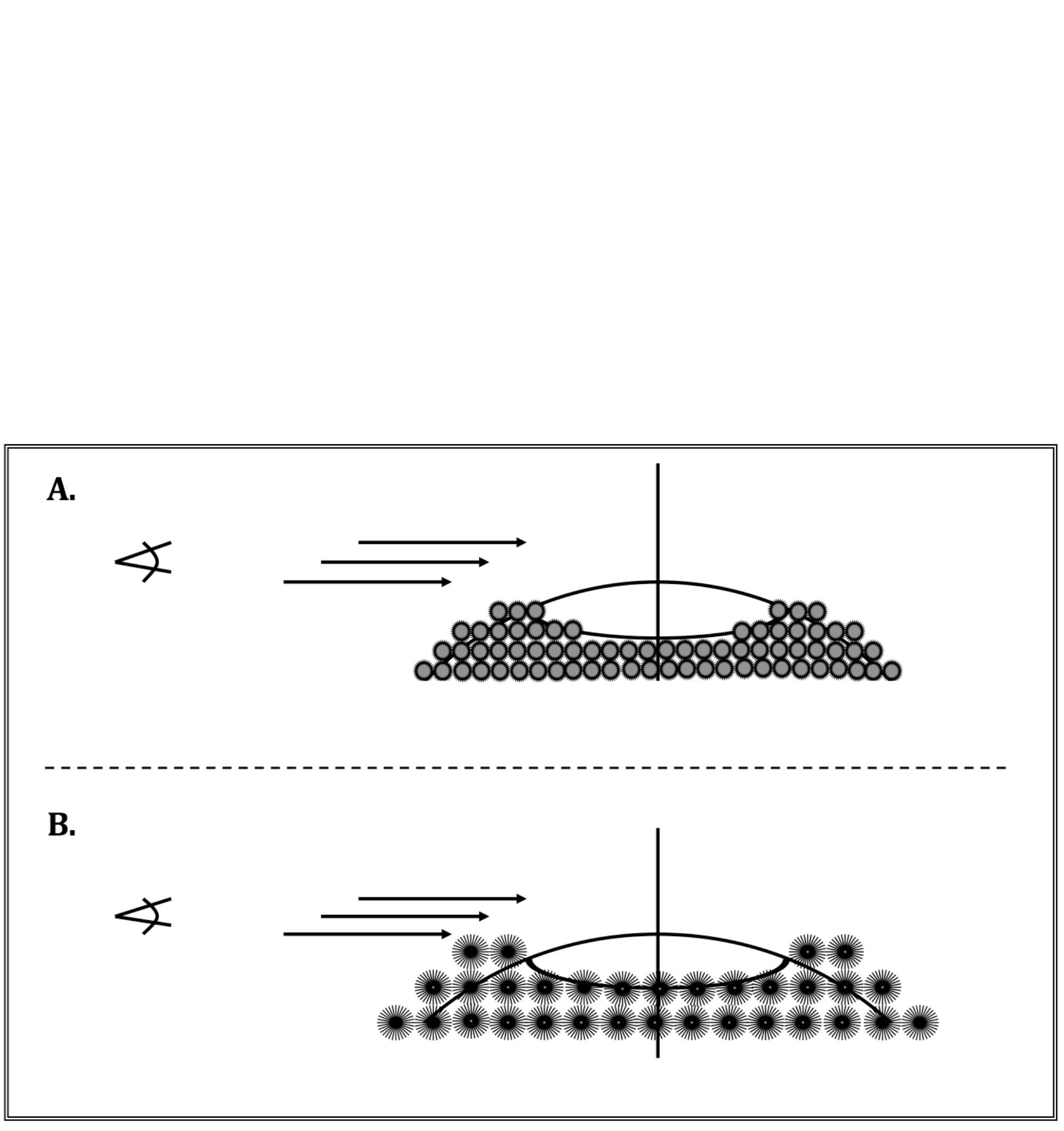}
              }
               \vspace{0.02\textwidth}   
     
\caption{A diagram showing the effects of the emitting plasma's scale height on the visibility of the edge of the polar coronal hole. The view shows a slice of the northern hemisphere of the sun at the central meridian; the rotation axis is also marked.  A) shows an example of a relatively small scale height contrasted with a large scale height of B). Notice how the emitting plasma in B) obscures the edge of the polar hole.}

       \label{Diagram}
\end{figure}

\begin{figure}     
   \centerline{\hspace*{0.0\textwidth}
               \includegraphics[width=0.43\textwidth,clip=]{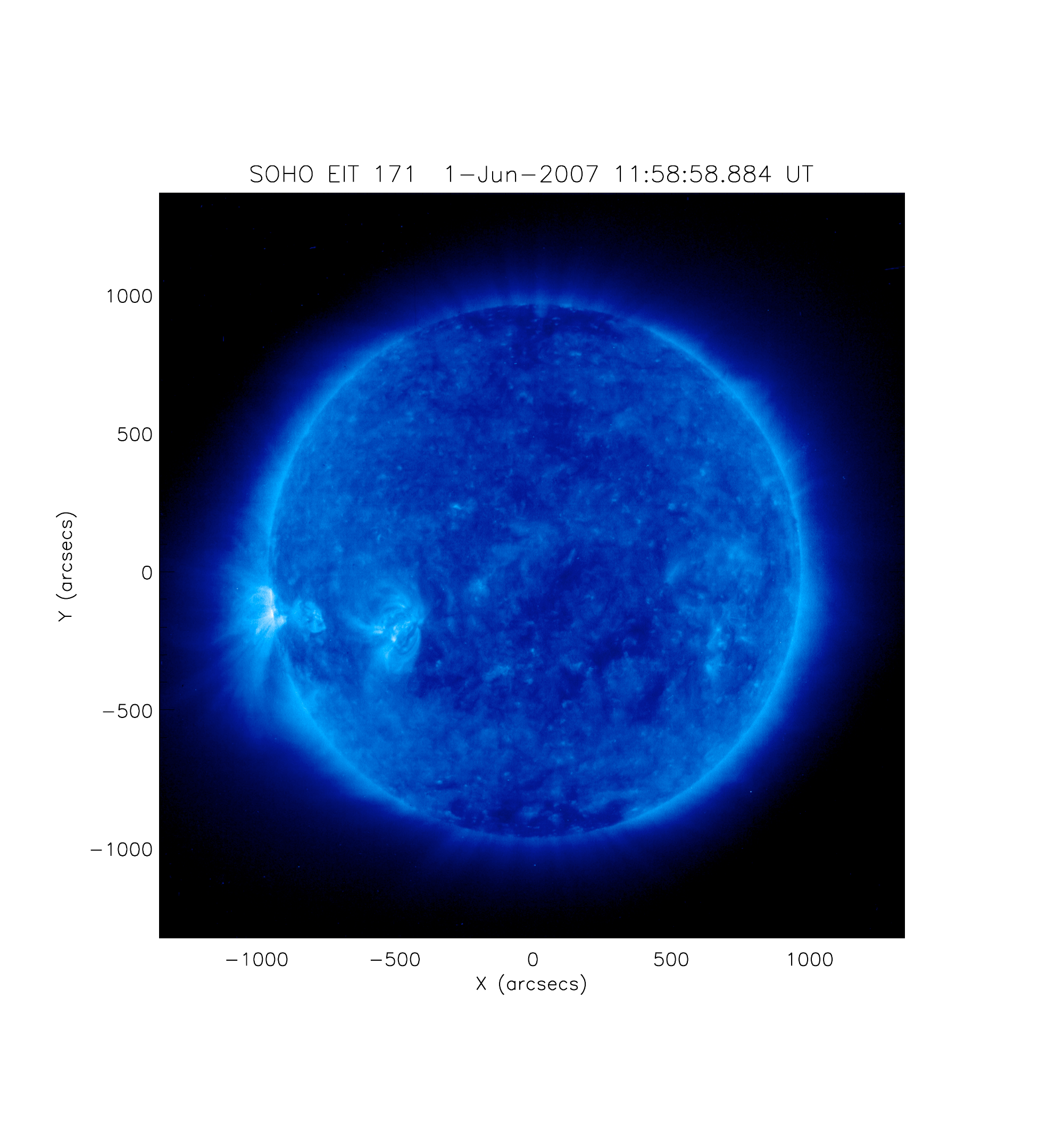}
               \hspace*{-0.1\textwidth}
              \includegraphics[width=0.43\textwidth,clip=]{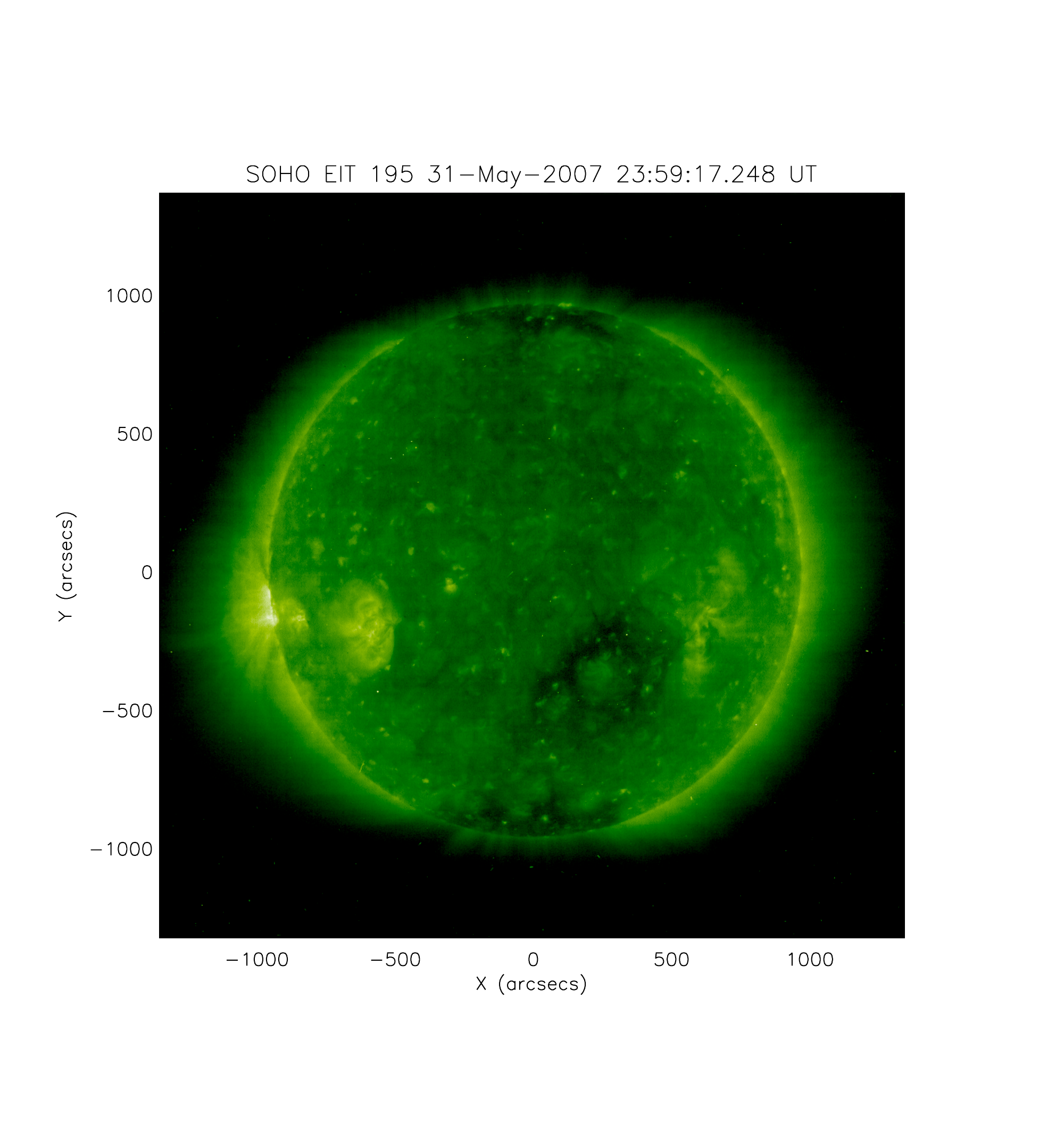}
              \hspace*{-0.1\textwidth}
               \includegraphics[width=0.43\textwidth,clip=]{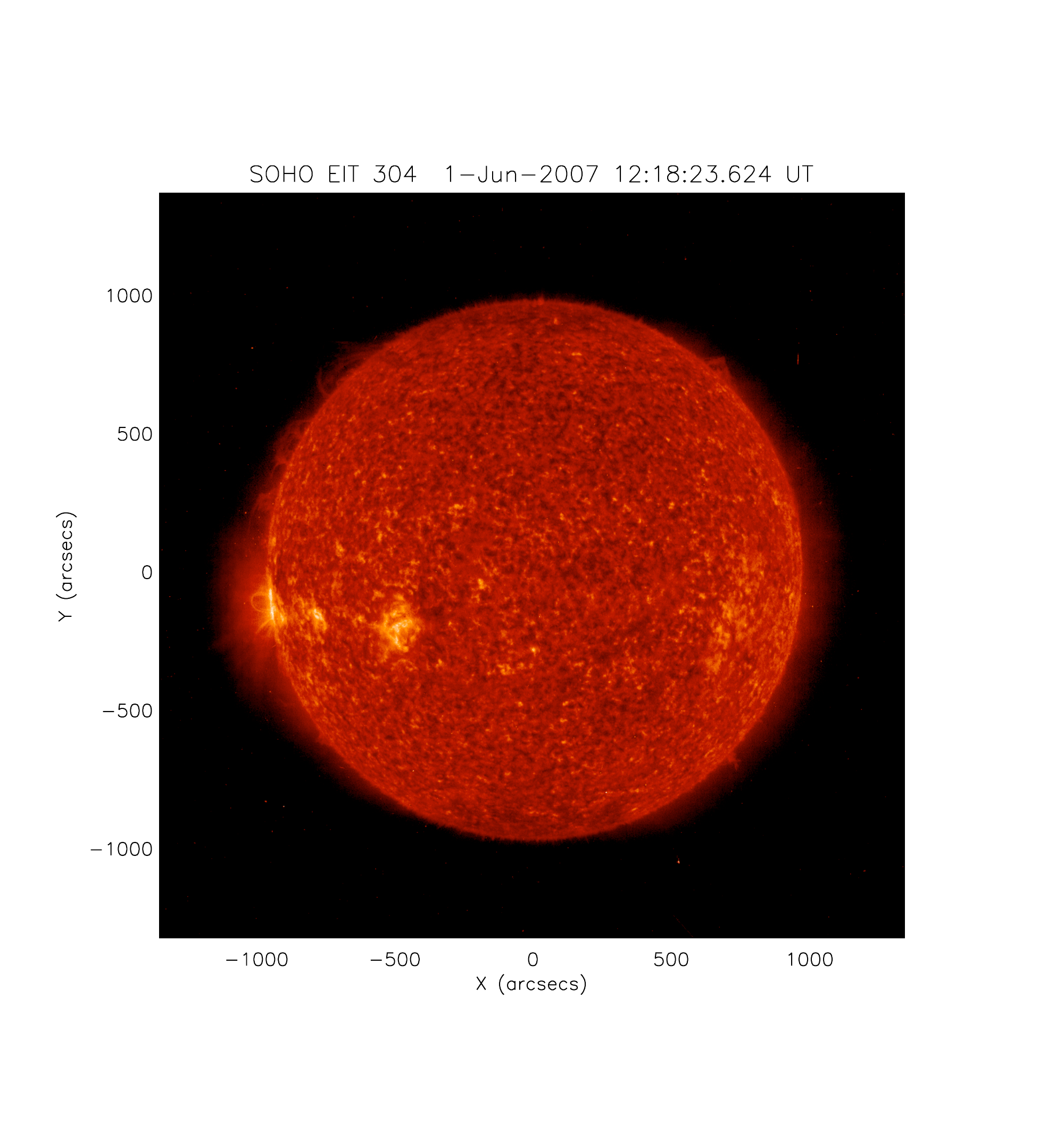}
                   }
               \vspace{-0.38\textwidth}   
     \centerline{ \bf     
      \hspace{-0.01 \textwidth}  \color{white}{A}
      \hspace{0.305\textwidth}  \color{white}{B}
      \hspace{0.3\textwidth}  \color{white}{C}
         \hfill}
     \vspace{0.3\textwidth}    

\caption{Examples of calibrated SOHO/EIT images at different wavelengths where polar coronal holes (PCHs) can be seen. A) 171~\AA, B) 195~\AA, and C) 304~\AA\ were all taken near June 1, 2007. The PCHs in these wavelengths appear as dark polar caps in both the north and south. The sizes and shapes of the PCHs vary between each wavelength due to differing scale heights in the emitting plasma.}

   \label{EIT-Images}
   \end{figure}

The appearance of coronal holes differs in each wavelength, which makes it unreliable to use one wavelength alone to measure them. Solar and Heliospheric Observatory's (SOHO) Extreme ultraviolet Imaging Telescope (EIT) observations over the last solar cycle gives us a unique opportunity to have extreme ultraviolet images of the sun in four different wavelengths at nearly the same moment in time \cite{Moses97}. EUV Fe lines measured by EIT have good contrast, but the emitting material has a relatively large scale height; for example, 195~\AA\ Fe~{\sc xii} has an emission that peaks at a temperature of $1.6 \times 10^6$ K and has an associated scale height between $60$ and $190$ Mm  \cite{Moses97}. When looking at the polar region on the solar sphere, if the emitting plasma has a small scale height, the holes are more easily discerned. However if the emitting plasma has a relatively large scale height, the edge of the polar hole is more easily seen at the limb of the the sun. In effect, the large scale-height is obscuring the edge of the polar hole at the central meridian (Figure~\ref{Diagram}). 
This would naturally lead one to choose a wavelength with the smallest scale height to get the most accurate measurement of the perimeter of the PCH. The He~{\sc ii} 304~\AA\ line has a small scale height compared to the other lines that EIT measures (He~{\sc ii} 304~\AA\ has an emission that peaks at a temperature of $8.0 \times 10^4$ K and has scale height between $3$ and $10$ Mm). However, the corrugations in the 304~\AA\ chromospheric network have a similar contrast to coronal holes, which makes discerning the boundary of the holes from the quiet sun network difficult. The high-temperature Fe~{\sc xv} 284~\AA\ images from EIT are particularly problematic to analyze, because at solar minimum, hot plasma is absent, resulting in poorly defined 284~\AA\ images \cite{Kahler02}. Thus, despite their limitations,  the Fe lines at 171 and 195~\AA\ and the He~{\sc ii} line at 304~\AA\ images are the best EUV images to use (Figure~\ref{EIT-Images}). 
The He~{\sc ii} 10830~\AA\ images from Kitt Peak could be used to supplement the EIT images; however they are left out of this survey to simplify the discussion. 

Because of the differential rotation rates of the sun, it would be inaccurate to use the Carrington rotation rate as the polar rotation rate. The mean solar rotation rate, $\omega$ (in degrees per day), is a function of heliographic latitude:
\begin{equation}
\label{Rotation-Rate}
\omega=A+B\sin^2\phi+C\sin^4\phi,
\end{equation}
where $A, B$ and $C$ are empirically determined coefficients and $\phi$ is the latitude. The values of the coefficients used are listed in Table~\ref{T-Rotation} \cite{Allens}. 
Coronal features, surface plasma and magnetic field features were used in determining the rotation rates of PCHs, because properties of coronal holes can fit into all three. The average rotation rate of these features is 10.88\degree\ per day or just slightly over 33 days for one rotation at a latitude of 70\degree. Thus we approximated the PCH rotation rate to be 33 days, from here forward called the Harvey rotation rate. 
For this work, we have defined the Harvey Rotation number (HR) to be
\begin{equation}
\mathrm{HR}=\frac{\mathrm{JD}-{2415024.5}}{33} ,
\end{equation}
where JD is the Julian day and Harvey rotation zero began 4 January 1900 (the beginning of Carrington Rotation 619).
Similarly, heliographic coordinates can be translated into an HR adjusted coordinate system by letting 0\degree\ in the HR coordinate system be the central meridian of the sun at the beginning of the Harvey rotation. 


\section{Detection Method}
	\label{S-Detection}
To determine the size of the polar coronal holes on sun using perimeter tracing, while avoiding some of the obscuration of the hole boundary when working in the polar region, the method of Henney and Harvey (2005) is adapted to a limb geometry. One image per day, as close to 0:00 UT as possible, is selected for each of the three EUV wavelengths. 

\subsection{Calibration}
	 \label{S-Calibration}
The images are first calibrated with the {\sf eit\_prep} software using the calibration tables available in the Solarsoft suite (Figure~\ref{EIT-Images}).
 They are also truncated so that the raw data numbers are whole numbers that fall between 0 and 1000. Because coronal holes have low emission in the EUV, this truncation does not affect the detection of the hole but does reduce the storage requirements of the image array. 

\subsection{Image Analysis}
	\label{S-Image Analysis} 

The image is then spatially smoothed. The morphological image analysis (MIA) functions Dilate ($\mathcal{D}$) and Erode ($\mathcal{E}$) are applied in series to remove small noisy structures. In general, MIA functions can be thought of as a 2D convolution of the shape operator and the original image.  Erode and Dilate are defined as:
\begin{equation}
\mathcal{D}(I,S)(i,j) \equiv \max_{<k,l>_S} \left[I(i-k, j-l) + S(k,l)\right]
\end{equation}
and
\begin{equation}
\mathcal{E}(I,S)(i,j) \equiv \min_{<k,l>_S} \left[I(i+k, j+l) - S(k,l)\right]
\end{equation}
where $I(i,j)$ is the original image array and $S(k,l)$ is a shape-operator array. The shape-operator array is defined by its displacement, $<\!\!k, l\!\!>_S$, from its origin, $(0,0)$  \cite{Michielsen01}.  

Erode and Dilate can be combined to create two other MIA functions Open ($\mathcal{O}$) and Close ($\mathcal{C}$), which are defined as:
\begin{equation}
\mathcal{O}(I,S)(i,j) \equiv \mathcal{D}\left[\mathcal{E}(I,S),S\right]
\end{equation}
and
\begin{equation}
\mathcal{C}(I,S)(i,j) \equiv \mathcal{E}\left[\mathcal{D}(I,S),S\right].
\end{equation}
The functions Open and Close are also idempotent, $\mathcal{O}(\mathcal{O}(I,S),S)= \mathcal{O}(I,S)$ and $\mathcal{C}(\mathcal{C}(I,S),S)=\mathcal{C}(I,S)$, a property that is advantageous in scientific image analysis. Generally, the Close function fills in small gaps between large regions of the same intensity. This has the effect of smoothing over regions that are too small to be part of a coronal hole while leaving larger regions largely intact. Features that do not fit the shape-operator ($S$) passed through the image are removed. The Open function, on the other hand, associates small features in close proximity to each other effectively grouping small regions of the same intensity into one large region. On solar images, using the Open function creates smooth edges between features. For a further discussion of MIA functions and their properties, see Michielsen and Raedt (2001).

In this case, we used a filter ($\mathcal{F}$) having a constant shape-operator ($S$) of a circular kernel with a radius of 3 pixels. The Filter is defined by: 
\begin{equation}
\mathcal{F}(I,S)(i,j) \equiv \mathcal{C}\left[\mathcal{O}(I,S),S\right],
\end{equation}
and has the effect of smoothing out the subtle variations in the image while leaving larger image features intact (Figure~\ref{Analysis}A). 
These image manipulations are needed to eliminate bright points that can occur in coronal holes and distract the hole boundary detection. It is also needed to blur out the grain structure in 304~\AA\ images and reveal the hole boundary.

\subsection{Thresholding}
	\label{S-Thresholding} 
The sun's intensity changes throughout the solar cycle. When detecting PCHs, this can be a problem, because the same intensity value may be a maximum in one image and a minimum in a different image of the same wavelength. To overcome this problem, an integrated intensity histogram of the remaining image is created. The integrated histogram is then divided by the total number of pixels in the image. From this histogram, a threshold level is picked to reveal the candidate holes in the disk (Figure~\ref{Analysis}B). 
The threshold level is different for each wavelength. For example, in a 171~\AA\ image, the threshold is the intensity-bin whose value is 0.70, or where 30$\%$\ of the pixels are darker than that intensity-bin. In a 195~\AA\ image, the bin value is 0.73; and in a 304~\AA\ image it is 0.50. These bin values were derived empirically to best expose the polar holes in each of the wavelengths. A binary image is then produced using the threshold level determined by the integrated intensity histograms where values less than the threshold is set to zero and all others are set to 1. 

Notice that, in the binary threshold image produced, exposing the polar holes on the limb of the sun overexposes holes near the disk center. The thresholding values distinguish this process from an equatorial detection method, because we set the threshold to specifically fit our limb geometry. Since we are only concerned with the limb, overexposing the equatorial holes is not a concern. By changing the threshold values we could easily adapt this algorithm to detect coronal holes in other regions. 

\subsection{Extraction} 
	\label{S-Extraction} 
\begin{figure}      
   \centerline{\hspace*{0.0\textwidth}
               \includegraphics[width=0.68\textwidth,clip=]{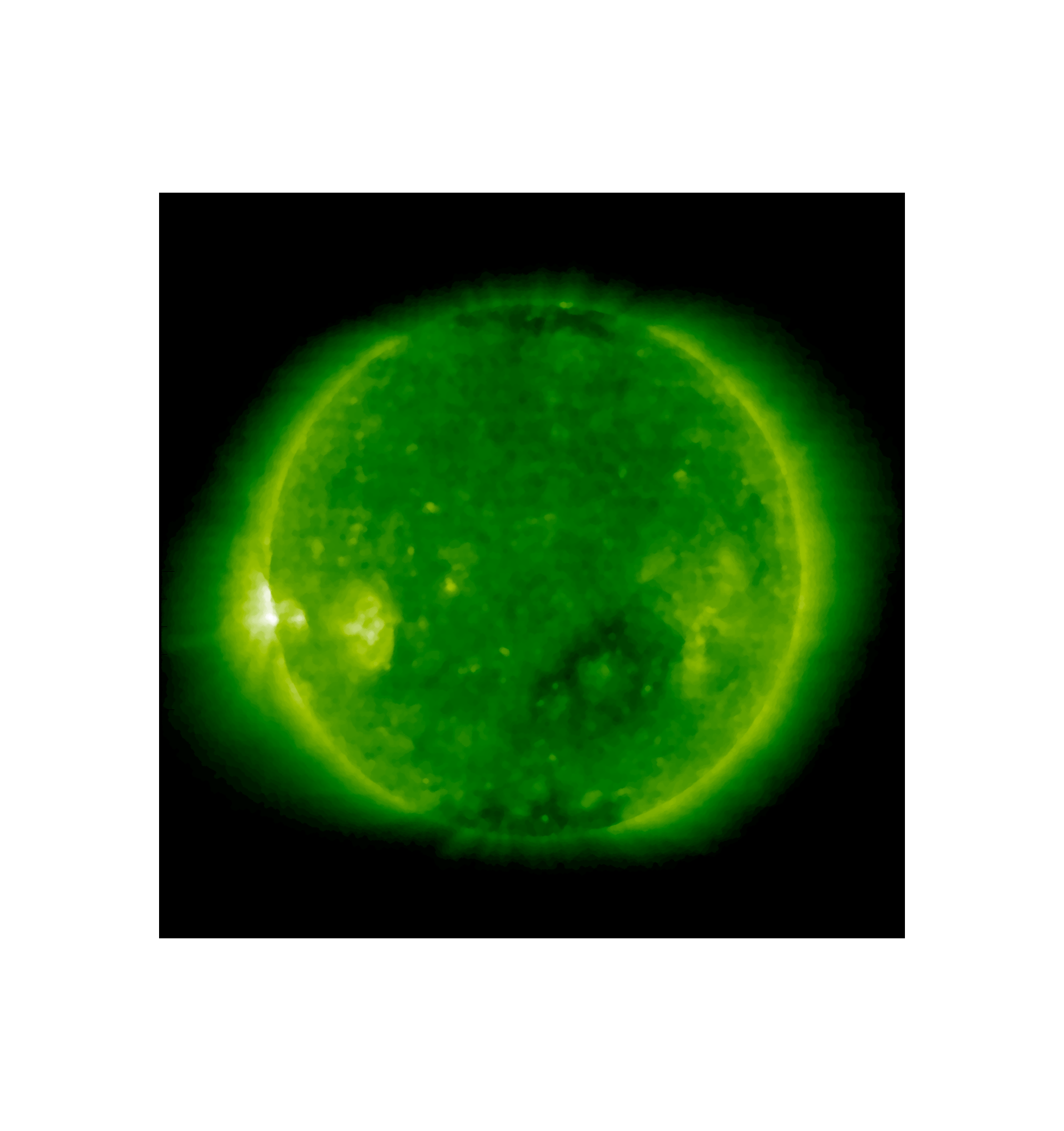}
               \hspace*{-0.18\textwidth}
              \includegraphics[width=0.68\textwidth,clip=]{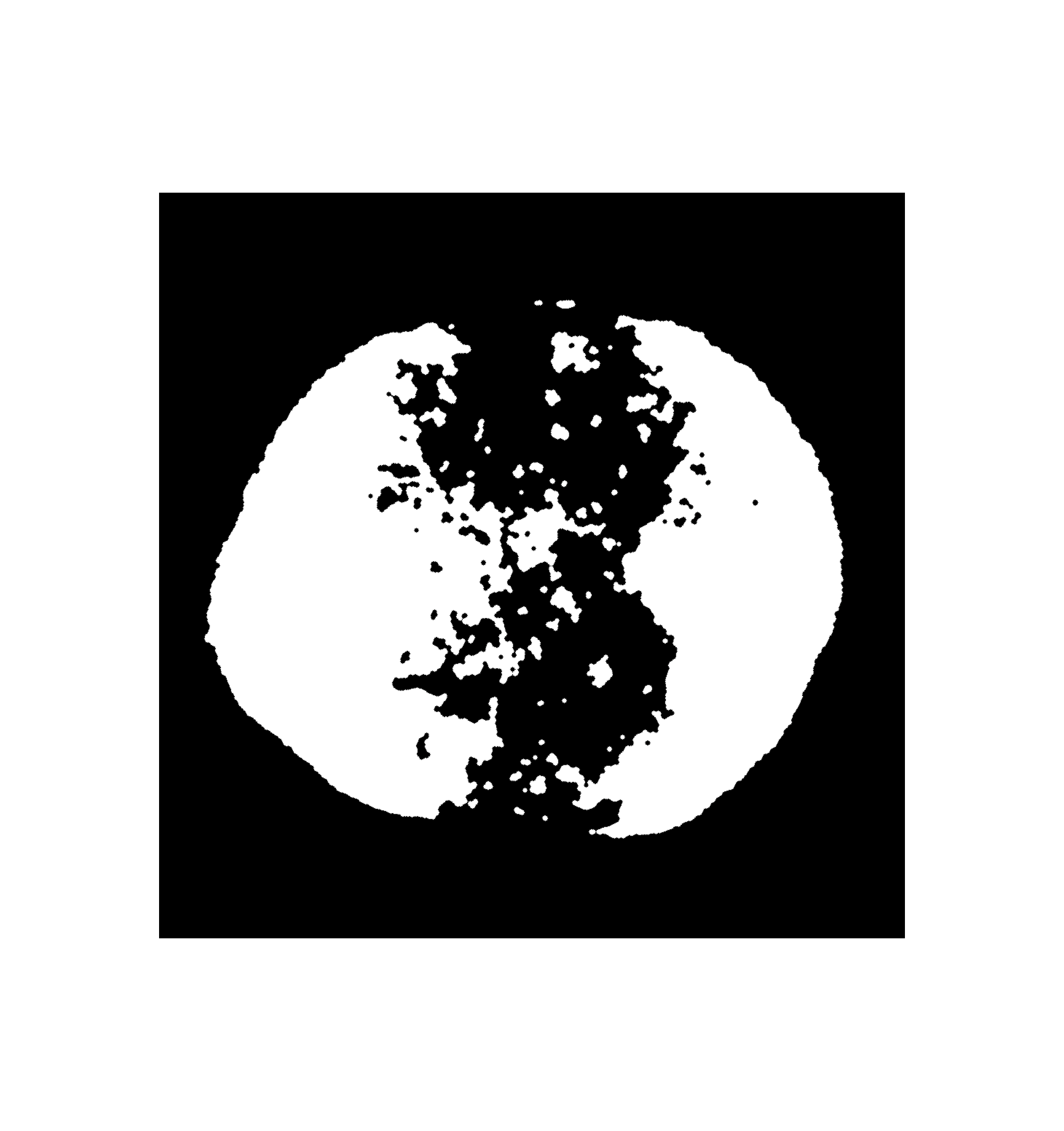}
              }
               \vspace{-0.59\textwidth}   
     \centerline{ \bf     
      \hspace{0.0 \textwidth}  \color{white}{A}
      \hspace{0.465\textwidth}  \color{white}{B}
         \hfill}
     \vspace{0.35\textwidth}    
   \centerline{\hspace*{0.0\textwidth}
               \includegraphics[width=0.68\textwidth,clip=]{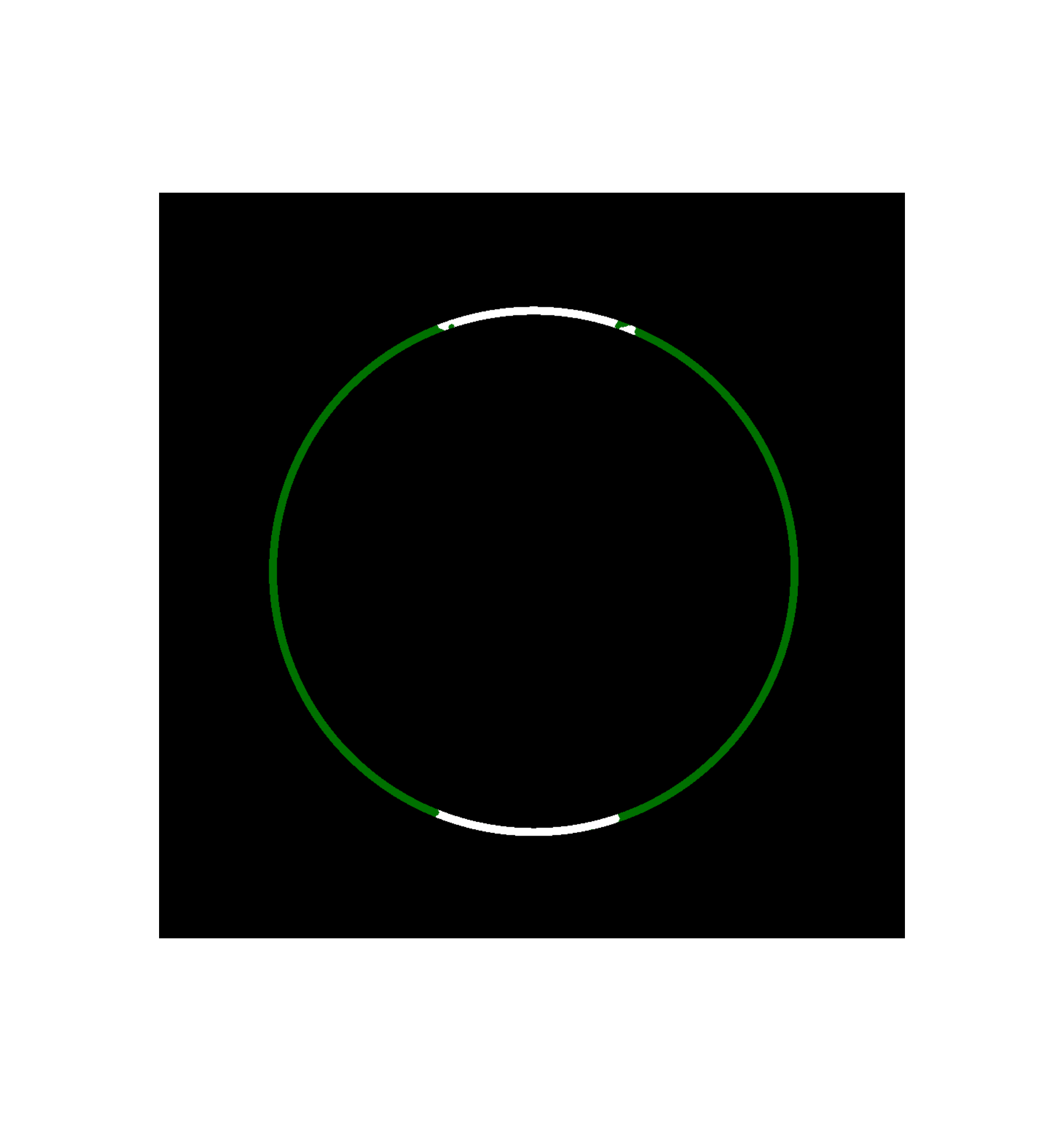}
               \hspace*{-0.18\textwidth}
               \includegraphics[width=0.68\textwidth,clip=]{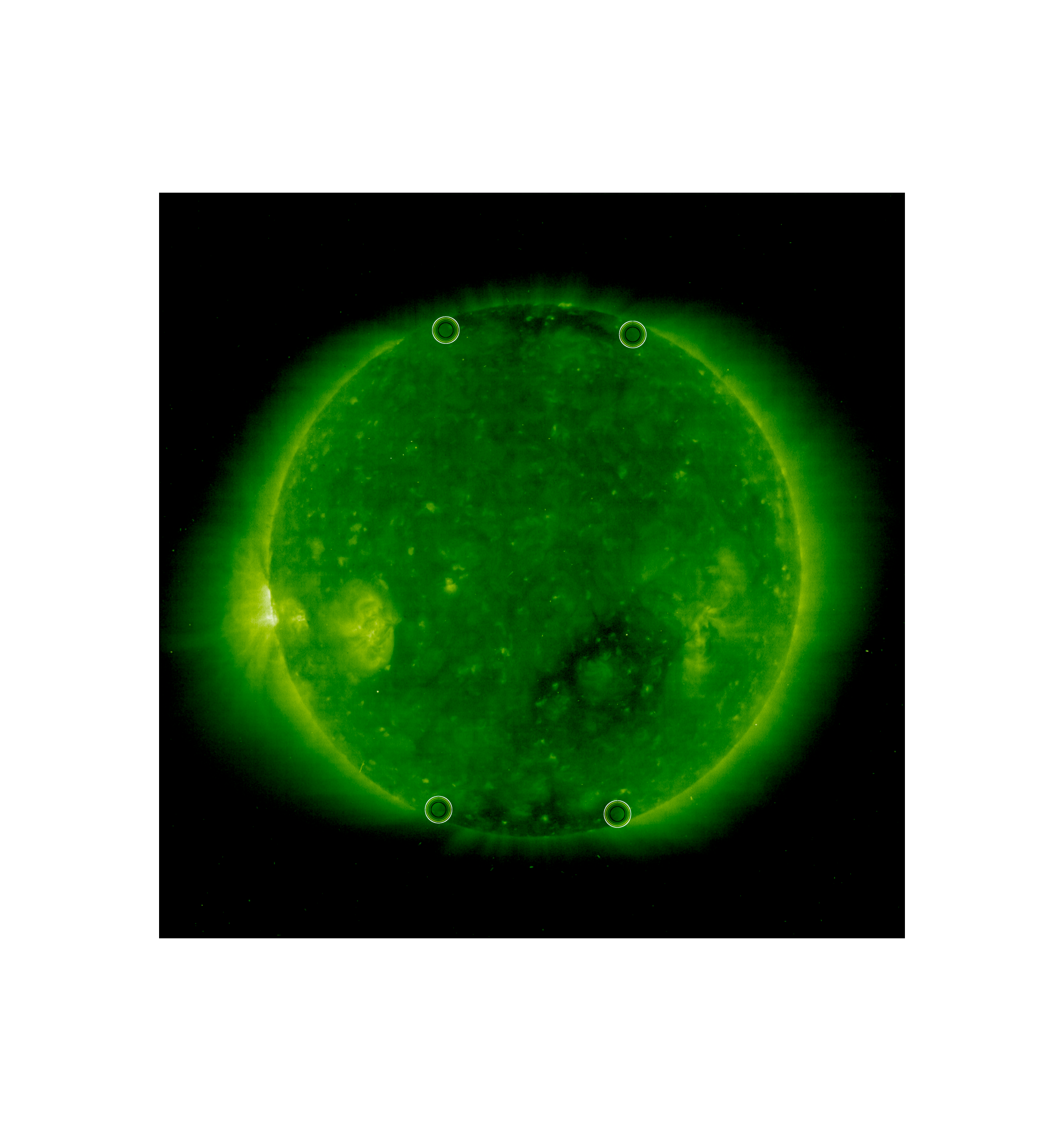}
              }
     \vspace{-0.59\textwidth}   
     \centerline{\bf     
      \hspace{0.01 \textwidth} \color{white}{C}
      \hspace{0.465\textwidth}  \color{white}{D}
         \hfill}
     \vspace{0.45\textwidth}    
              
\caption{An illustration of the automated PCH detection process using the 195~\AA\ image shown in Figure ~\ref{EIT-Images}: A) After using morphological transform functions Close and Open to blur the original image; B) A binary image of the image shown in A) retaining the top 73$\%$ of the bin values in an integrated intensity histogram; C) An annulus of the outter 6\% of solar disk after removing off-limb and central disk data; D) The edges of the north and south PCHs are shown with a circle. The heliographic coordinates are then calculated and used to mark the perimeter of the PCH after translation to a Harvey coordinate system.}

       \label{Analysis}
   \end{figure}

 Next, we isolate the disk itself by trimming the coronal streamers outside the limb of the sun. The location of the solar limb is determined using limb brightening as defined by the EIT package in SolarSoft. The center of the disk is also trimmed, leaving an annulus spanning the outer 6$\%$\ of the solar radius. Then the polar regions are isolated by masking the disk up to 60\degree\ north and south latitude (Figure~\ref{Analysis}C). 
The location of the PCH boundary at the limb can now be extracted. 
 
\begin{figure}      
   \centerline{\hspace*{-0.0\textwidth}
               \includegraphics[width=0.47\textwidth,clip=]{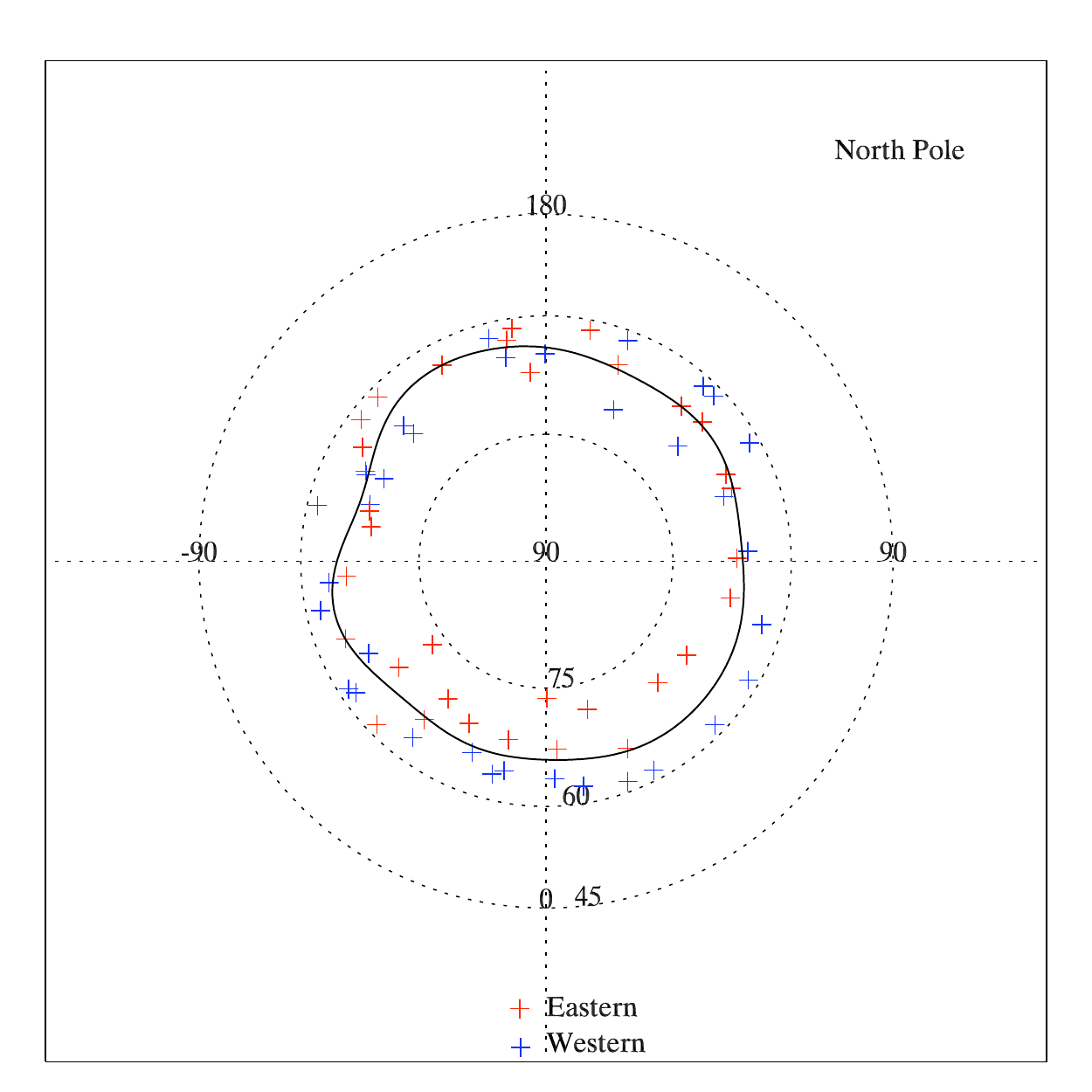}
               \hspace*{-0.0\textwidth}
              \includegraphics[width=0.47\textwidth,clip=]{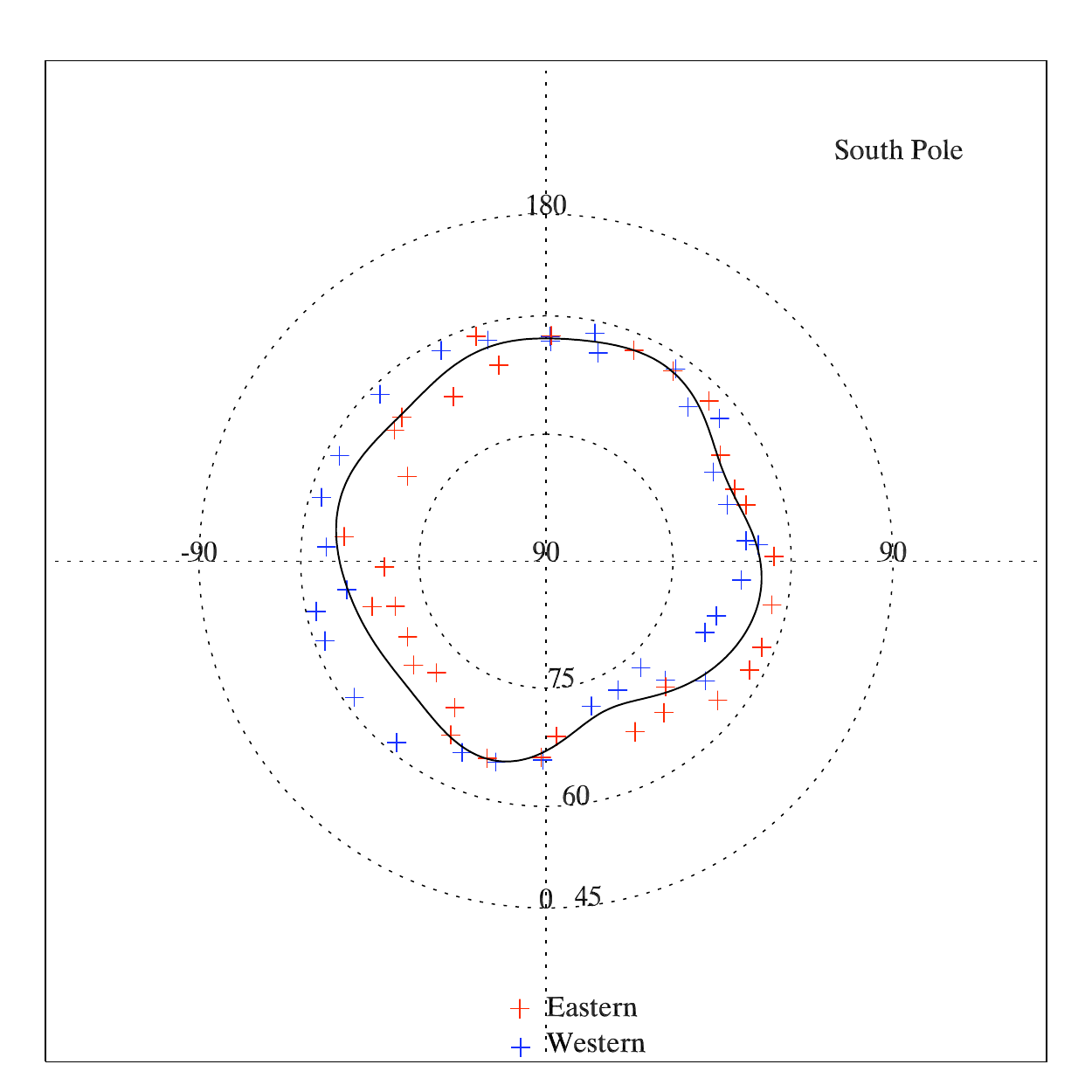}
              }
     \vspace{0.0\textwidth}    

\caption{The north and south pole limb measurements in Harvey coordinates beginning 12 January 2006.  Each blue and red mark designates the coordinates of the edge of the PCH on the ascending and descending limb respectively. A fit to the data using Equation~(\ref{fit-equ}) with $N_{\rm max}=7$ is drawn as the solid line. The center of the PCHs is coincident with the rotation axis. }

       \label{Fitting}
\end{figure}

Once the candidate hole edges are identified, their heliographic longitudes are calculated and translated into Harvey coordinates (Figure~\ref{Analysis}D). 
Using a Harvey coordinate system to determine the edge of the holes has the added benefit of allowing the perimeter of the hole to emerge organically from the data; {\it i.e.} the perimeter tracks back onto itself after a 33 day rotation. The points are then plotted on a polar grid using the HR rate
(Figure~\ref{Fitting}).

To evaluate the quality of the measurement, we find the fraction of pixels in the remaining binary limb within the candidate hole edge that appear as a hole ($Pixels_{\mathrm{zero}}$) as compared to all of the pixels (${\rm Pixels}_{\mathrm{total}}$) that fall within that candidate hole boundary, {\it i.e.}, within the boundaries identified in Figure~\ref{Analysis}D, 
the number of white pixels is compared to the total pixels in Figure~\ref{Analysis}C
 (eq.~\ref{E-quality}).  

\begin{equation}
	\label{E-quality}
	\mathrm{Quality} = \frac{\mathrm{Pixels}_{\mathrm{zero}}}{{\rm Pixels}_{\mathrm{total}}}
\end{equation}

If the hole is poorly defined, then the quality is low.  If the quality of the measurement of the hole boundary is less than 15\%\ then the measurements are discarded. This is to help guard against false positive measurements due to equatorial holes encroaching into the polar region, especially when the sun is active.

\subsection{Method Verification}
	\label{S-Verification}  
\begin{figure}      
   \centerline{\hspace*{-0.0\textwidth} 
               \includegraphics[width=0.47\textwidth,clip=]{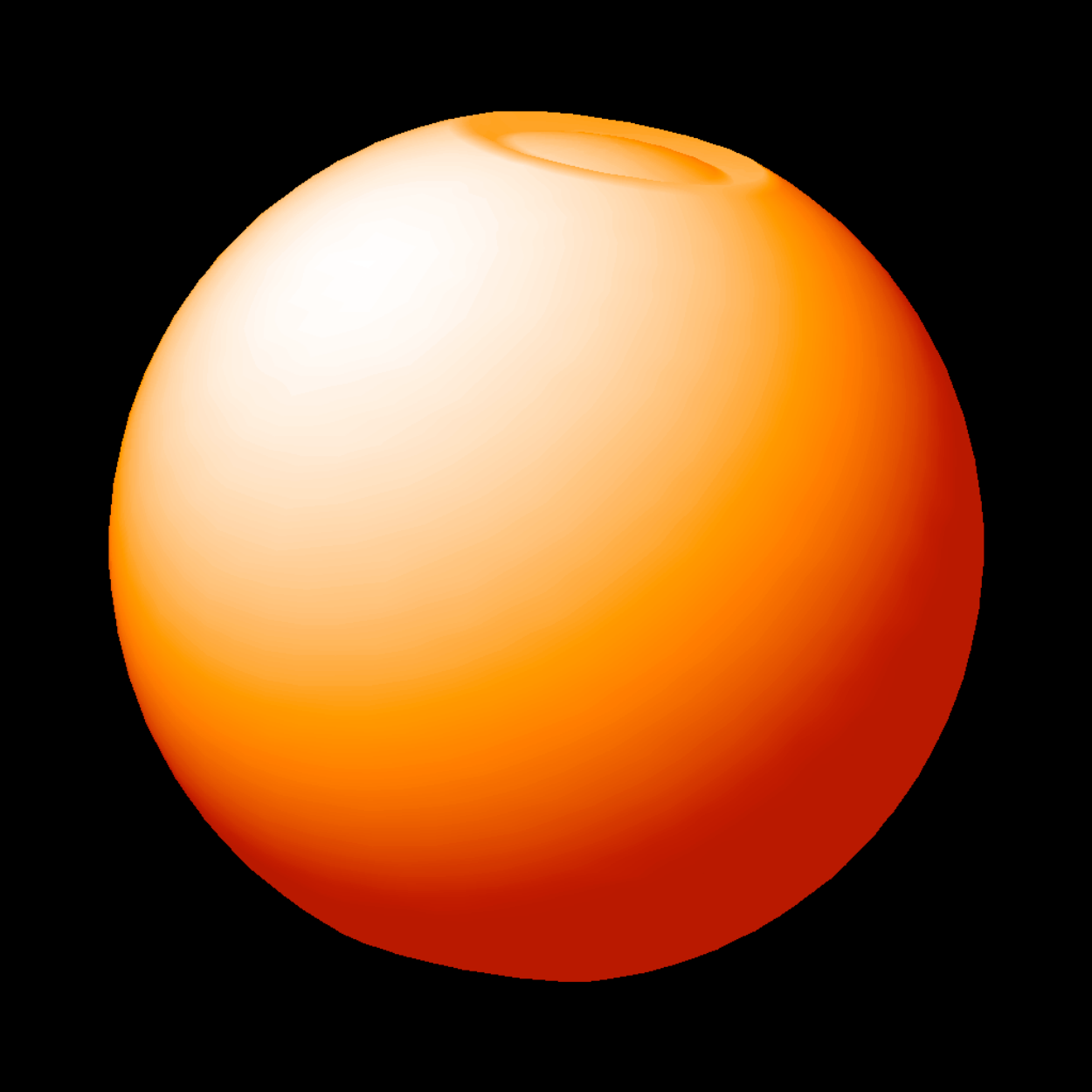}
               \hspace*{-0.0\textwidth}
              \includegraphics[width=0.47\textwidth,clip=]{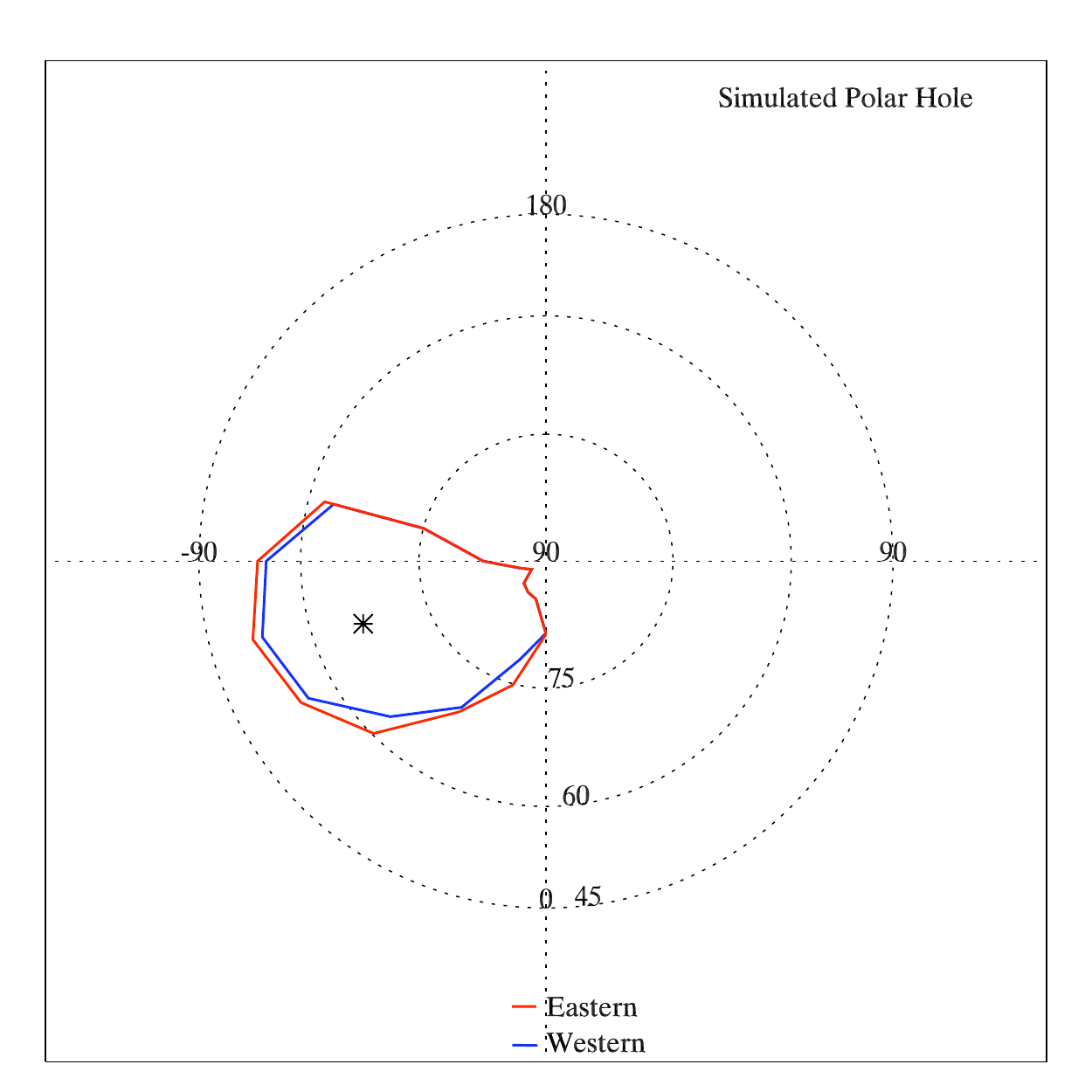}
              }
               \vspace{-0.45\textwidth}   
     \centerline{\bf     
      \hspace{0.035 \textwidth}  \color{white}{A}
      \hspace{0.44\textwidth}  \color{black}{B}
         \hfill}
     \vspace{0.45\textwidth}    

\caption{Example of converting a simulated PCH to a polar projection using limb detection and perimeter tracing. A) One of the images of the simulated sun with a northern PCH. The simulated hole has an offset of 20\degree\ from the pole and a diameter of 18\degree\ at its lowest point. The whole simulated sun has a heliographic tilt of -7\degree.  B) The perimeter found by the perimeter tracing technique described in Section~\ref{S-Detection}. The PCH at the ascending (western -- blue) and descending (eastern -- red) limb of the sphere are shown. The detected hole has a diameter of 20\degree, and its center (marked with a star) has an offset of 23\degree.  }

       \label{Testing}
\end{figure}

To test the perimeter tracing algorithm we generated a pseudo-coronal hole on a computer-generated sun. The circular hole was 20\degree\ offset from the rotational axis, 18\degree\ in diameter beveled up to the surface (the hole is about 23\degree\ in diameter at the surface), and the rotational axis was inclined -7\degree\ from our visual vertical axis (simulating the most extreme heliographic tilt seen during the year, Figure~\ref{Testing}A). 
We rotated the model through 360\degree\ in 15\degree\ steps letting the detection program find the edges of the simulated hole every step. Then, after assigning spherical coordinates to each of the detected points, we mapped them to a polar grid (Figure~\ref{Testing}B). The result returned a hole centered at a colatitude of 23\degree\ with a diameter of 20\degree. The diameter is not exactly 18\degree\ because the detection method is sensitive to the bevel of the simulated hole. Also, the detected limb of the simulated hole is not exactly circular because of the projection effects as the hole sweeps around.


\section{Data Analysis}
	\label{S-Data} 
When looking at an HR of polar measurements, it would be imprudent, although tempting, to simply connect the data points and say that this represents a good measurement of the PCH boundary; however, it would tend to emphasize the effects of the noise in the measured data and would produce a less accurate area bounded by the limb measurements of the hole perimeter. In Section~\ref{S-Detection} we accounted for this by using morphological transforms to smooth the candidate hole boundaries. When assessing the area within the coronal hole boundaries, fitting the collected data with a line of best fit keeps the integrity of the physical nature of coronal holes.

\subsection{Functional Fitting}
	\label{S-Fitting}
	
The first step in the fitting process is to combine the two separate limb measurements into one data set. This is where translating into Harvey coordinates in Section~\ref{S-Extraction} pays off. Combining the sets of points is as simple as creating a large array from the two smaller arrays, since Harvey coordinates are the appropriate reference frame for polar features.
 
A first order filter is then applied. If greater than half of the measured points are missing in a given Harvey rotation, the entire rotation is thrown out. This filter is needed to guard against under-sampling the limb measurements of the PCH.
  
The array is then translated so that the center of mass of the points is the origin of the coordinate grid. This translation is necessary to force the fitting routine to fit each point more efficiently. For example, if the polar hole intersected the origin, the fitting routine would tend to extend into negative space to fit the feature, creating an unrealistic fit of the hole. Translating the PCH measurements provides some protection against this circumstance.
   
The array of PCH measurements can then be fitted with a Fourier cosine series in the longitude: 
\begin{equation}\label{fit-equ}
	f(\phi)=a_{0}+\sum_{n=0}^{N_{\mathrm{max}}} a_{2n-1}\cos(n\phi+a_{2n}).
\end{equation}
	$N_{\mathrm{max}}$ is allowed to vary from two to seven to generate several candidate fits. 
   
The points are fitted using a non-linear least squares fitting routine with an arbitrary number of input parameters. This method uses an iterative gradient-expansion algorithm combined with a minimization of $\chi^2$ ~\cite{Numerical_Reci}. A full discussion of the fitting procedure can be found in Bevington and Robinson (1992). A similar method was used by \cite{Pesnell2000} to analyze ozone variations in the Earth's atmosphere.
   
To increase the viability of the fit produced by the procedure described above, weights are assigned to each measurement.  Because PCHs do not have discontinuities, we can say adjacent measurements are more accurate if they do not have large variations.  In order to give greater weight to points that are more closely associated while putting less emphasis on outlying points, a Gaussian weighting system is used. The total array is divided up into octants: between 0\degree\ and 45\degree, 45\degree\ and 90\degree, {\it etc}.  The variance, $\sigma^2$, is then calculated within each octant, for the angular distance to the pole ($\theta$) where the standard deviation in the radius is~\cite{Bevington92}:
\begin{equation}\label{stdev}
	\sigma_i = \sqrt{\frac{1}{N-1} \sum_{i=1}^{N}\left(\theta_i - \bar{\theta}\right)^{2}} .
\end{equation}
Since points that have a small radial variance mean that the edge of the PCH is better defined, we can weight the points as the reciprocal of the the variance in each octant. Note that all the points in an octant are given the same weighting. 
   
Since we generate six different lines of best fit using Equation~(\ref{fit-equ}), we must make some judgments about which fits are better than others. If the fitting line ($f[\phi]$) extends below 45\degree\ latitude, the line is no longer fitting in the polar region and we can disregard that fit. Next, we must be careful of over-fitting the given polar measurements. A symptom of over-fitting is a large length of the curve ({\it i.e.}, it has a lot of wiggles) that defines the perimeter of the PCH. The arc length of the fitted curve is found analytically by integrating the generated line of best fit ($f[\phi]$). If a fit has a total length that is greater than four times the smallest curve's length, we disregard the larger fit. 
   
Of the remaining lines of best fit for the given PCH data, we can now find the area. The area of the polygon formed by the line of best fit can be found by integrating $f(\phi)$ from Equation~(\ref{fit-equ}) \cite{Beyer91}.  Another affect of over-fitting a series of points is that a larger than necessary area is enclosed. Thus we can eliminate all fits that are greater than twice the smallest enclosed area.
    
From the fits that remain, taking a simple average of the areas they enclose gives us a good sense of the fitted area of the PCH (the black line in Figure~\ref{Fitting}). 

\subsection{Statistical Analysis}
	\label{S-Stats} 
	
It is important to define a confidence measurement when working with fits to a set of data. To accomplish this, we estimate the probable uncertainties in the fitted parameters. Because each of the data points is independent, each contributes its own bit of uncertainty to the fitting parameters. Propagating these errors, it can be shown that the variance $\sigma^{2}_{f}$ in the value of the function will be
\begin{equation}
\sigma_f^2=\sum_{i=1}^{N}\sigma_i^2\left(\frac{\partial f}{\partial  y_i}\right)^2.
\end{equation}
 Directly evaluating the partials of each of the fitting parameters from the solution will produce the variances in each of the parameters~\cite{Numerical_Reci}.

With the variances in each parameter, we can then produce a one-sigma confidence range.  By subtracting one sigma from each of the parameters, a minimum area fit is produced. Likewise, by adding one sigma, a maximum area fit can also be determined. However, since the fitting function is a transcendental function, the error range does not always produce realistic results. When the fitting function has a tenuous relationship with a few outlying points, adding or subtracting a one-sigma error can cause significant ringing in the fitting function leading to an unrealistic fit. This problem is most evident during solar maximum where the measured points are not well correlated. 

To determine the range of values seen in Figure~\ref{Results},
 we found the maximum and minimum areas associated with each of the fits determined to be good in Section~\ref{S-Fitting}. Using the same method as used with finding the area bounded by a line of best fit, we averaged the maximum values and minimum values respectively. This produced the one-sigma range of values. Where the errors were not realistic, the area bounded by the best fit was plotted. 

\section{Results and Discussion}
	\label{S-Discussion} 
\begin{figure}      
   
   \centerline{\hspace*{-0.0\textwidth}
               \includegraphics[width=1.4\textwidth,angle=90]{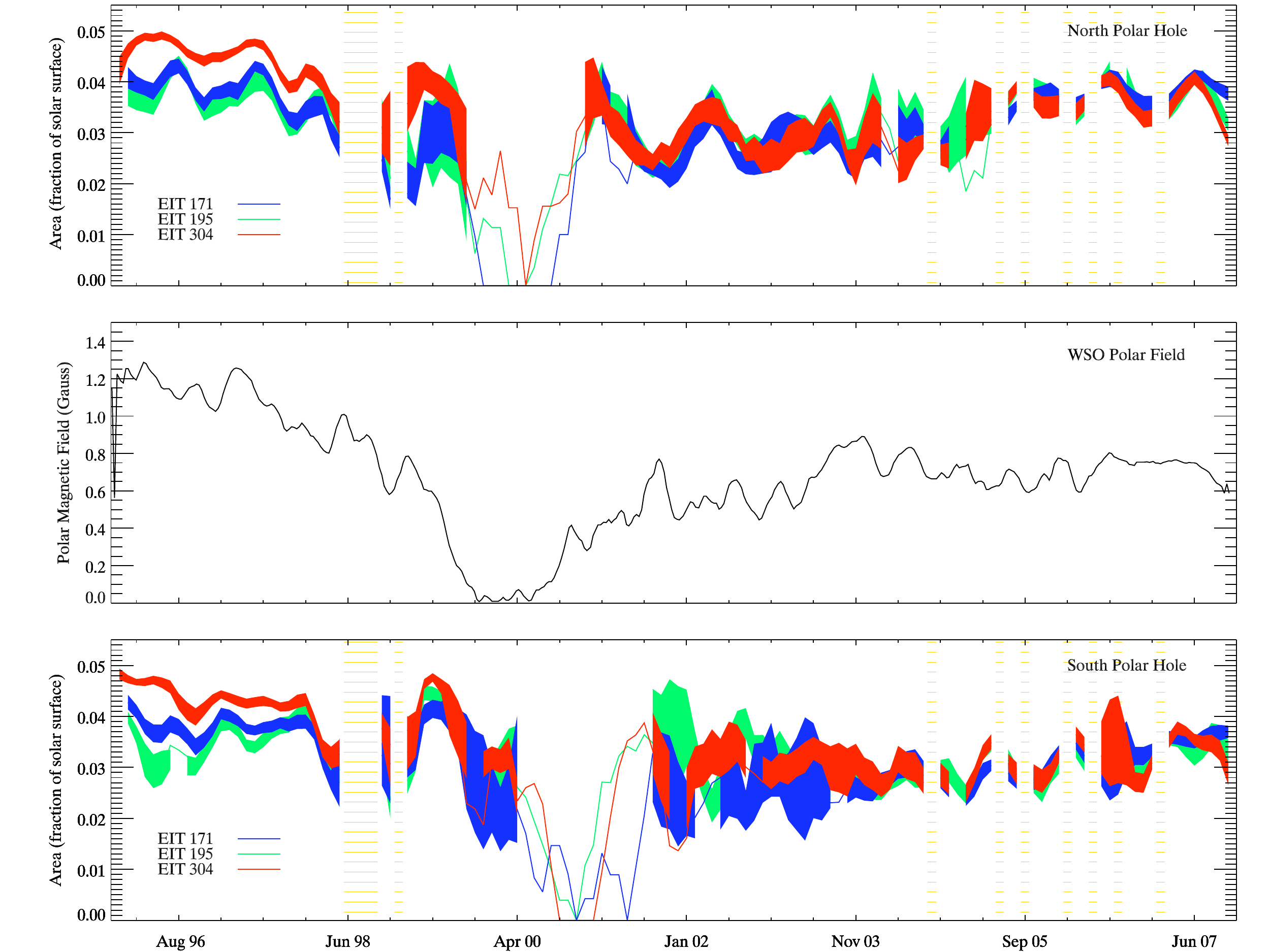}
              }
               \vspace{0.02\textwidth}   

\caption{A time series of PCH areas as a percent of the total solar surface area. Each area is calculated over a 33-day Harvey rotation using the perimeter tracing technique. The blue line corresponds to 171~\AA, the green to 195~\AA, and the red to 304~\AA. All lines are smoothed with a 3 point boxcar function. The yellow hash-marked sections on the graphs mark times when more than half of the Level 0 EIT data was missing for a given Harvey rotation. The absolute value of the averaged WSO polar magnetic field from Figure~\ref{WSO-figure} is included for reference.}

       \label{Results}
\end{figure}

Studying the evolution of the polar coronal holes over solar cycle 23 utilizes the entire lifetime of SOHO and chronicles the decline of the PCH into solar maximum and the rise to the current solar minimum (Figure~\ref{Results}). 
The northern PCH disappears in the final months of 1999 and returns in the last quarter of 2000. On the other hand, the southern PCH does not disappear until mid 2000 and does not return in earnest until the first quarter of 2002. This asymmetry is consistent with past surveys of PCHs.
 
The northern PCH is significantly better defined than the southern hole near solar maximum, as seen by the large error in the southern measurements in 2002 and 2003. The large error measurements in the south are confirmed by a qualitative assessment of the disk images during 2002 and 2003. These disk images appear to have large equatorial holes migrating toward the southern polar region making it difficult to quantify the boundary of the PCH. It is also interesting to note the relative disagreement of PCH area of the three wavelengths in 1996 and into 1997 in contrast to their relative agreement (within each otherÕs error estimates) for the rest of the series. 
   
Because of the relatively small time series of this study, it is difficult to make broad assertions about the nature of the polar holes. Nonetheless, comparing this series to previously compiled data reveals some noteworthy results.
   
Comparing this work to that of Harvey and Recely (2002) reveals some striking similarities. Harvey and Recely produced a time-series plot of the PCH area over the second half of solar cycle 22 and first half of 23. Harvey and Recely have the northern polar hole shrinking to less than 1\% of the surface area of the sun midway through 1999 and Harvey's ``pre-polar hole" reemerging at the beginning of 2001. They also show the southern PCH disappearing completely in mid 2000 but end their study before the southern hole reappears. Figure~\ref{Results} 
shows a similar time evolution of the northern and southern holes.
    
After the northern PCH reemerges, Harvey and Recely (2002) observed an area of about 4.8\% of the surface area of the sun. By comparison, this work observes a northern area of about 3.5\% or an He I 10830~\AA\ area that is about 37\% greater than the EUV-derived area. In 1996, Harvey and Recely report the northern PCH area to be about 8.25\% the solar surface area. We observe an area of about 4.2\%, meaning that the He I 10830~\AA\ area is almost twice the EUV-derived area at the beginning of solar cycle 23.
    
Another important comparison is with the Wilcox Solar Observatory (WSO) measurements of the polar magnetic field (Figure~\ref{Results}).  
The WSO reports a magnetic flux magnitude of about 1.0 gauss at the minimum preceding cycle 23, while it reports about 0.6 gauss at the current minimum. The polar field is 40\% less now than it was a cycle ago. Because coronal holes are characteristically threaded by open magnetic field lines and the solar polar magnetic fields are confined to relatively small caps~\cite{Sheeley89}, this seems imply that the size of the polar hole should be significantly smaller as well. However, we observe the size of the hole to be at most 15\% smaller this solar minimum as compared to last. This discrepancy would suggest that the polar coronal hole is more a product of dynamics rather than magnetic fields. 

When comparing the WSO time series of the magnetic field to the area of the polar holes at both poles, it appears that there is a close correlation between them. A correlation plot of the north or south hole and the WSO field reveals no more than a weak correlation between the plots seen in Figure~\ref{Results}. If we separate the average WSO field into its northern and southern components, the correlation seen is about the same as with the average magnetic field. 

\section{Comments and Conclusions}
	\label{S-Conclusions} 

We have shown that perimeter tracing on disk images liberates the analysis of PCHs from a reliance on synoptic maps. Perimeter tracing in the polar region, given the differential coronal rotation rate, demands the development of a new coordinate system rather than a mid-latitude Carrington rotational period. A Harvey coordinate system better represents the polar rotation rate and allows us to get a more accurate representation of the PCH. Using EIT images over solar cycle 23, we observe the PCH area of the north at the end of 2007 to be about 3.6\% of the total solar surface area versus about 4.2\% in 1996. The area of the southern PCH at the end of 2007 is observed to be about 3.4\% as compared to about 4.0\% in 1996. The estimate of the PCH size between 1996 and 1999 is complicated, because the three EIT wavelengths used have distinctly different areas. After the beginning of 1999 the areas of the different wavelengths agree more closely.   Yet even with this difference, we observe both the north and south PCH area to be no more than 15\% smaller now than they were at the beginning of cycle 23. 

It is possible to incorporate magnetograms into this PCH detection method by summing the net magnetic field within the derived hole boundary. Using magnetograms would allow us to measure the magnetic flux density evolution over the solar cycle and better isolate the polar hole from equatorial holes.  Perimeter tracing is also extendable into the past with archives of He I 10830~\AA\ images, which will allow us to get a better idea of the long-term evolution of PCHs.

\begin{acks}
 This work was supported by NASA's Solar Dynamics Observatory (SDO). The EIT images are courtesy of the SOHO/EIT consortium. The referenced SolarSoft software can be found at \textsf{www.lmsal.com/solarsoft/}.
\end{acks}

\mbox{}~\\
\bibliographystyle{spr-mp-sola-cnd}    
\bibliography{PCH_Draft_v3}  

\begin{thebibliography}{22}
\ifx \bisbn   \undefined \def \bisbn  #1{ISBN #1}\fi
\ifx \binits  \undefined \def \binits#1{#1} \fi
\ifx \bauthor  \undefined \def \bauthor#1{#1} \fi
\ifx \batitle  \undefined \def \batitle#1{#1} \fi
\ifx \bjtitle  \undefined \def \bjtitle#1{\textit{#1}}\fi
\ifx \bvolume  \undefined \def \bvolume#1{\textbf{#1}}\fi
\ifx \byear  \undefined \def \byear#1{#1} \fi
\ifx \bissue  \undefined \def \bissue#1{#1} \fi
\ifx \bfpage  \undefined \def \bfpage#1{#1} \fi
\ifx \blpage  \undefined \def \blpage #1{#1} \fi
\ifx \burl  \undefined \def \burl#1{\textsf{#1}} \fi
\ifx \doiurl  \undefined \def \doiurl#1{\textsf{#1}} \fi
\ifx \betal  \undefined \def \betal{\textit{et al.}} \fi
\ifx \binstitute  \undefined \def \binstitute#1{#1} \fi
\ifx \bctitle  \undefined \def \bctitle#1{#1} \fi
\ifx \beditor  \undefined \def \beditor#1{#1} \fi
\ifx \bpublisher  \undefined \def \bpublisher#1{#1} \fi
\ifx \bbtitle  \undefined \def \bbtitle#1{\textit{#1}} \fi
\ifx \bedition  \undefined \def \bedition#1{#1} \fi
\ifx \bseriesno  \undefined \def \bseriesno#1{\textbf{#1}} \fi
\ifx \blocation  \undefined \def \blocation#1{#1} \fi
\ifx \bsertitle  \undefined \def \bsertitle#1{\textit{#1}} \fi
\ifx \bsnm \undefined \def \bsnm#1{#1} \fi
\ifx \bsuffix \undefined \def \bsuffix#1{#1} \fi
\ifx \bparticle \undefined \def \bparticle#1{#1} \fi
\ifx \barticle \undefined \def \barticle#1{#1} \fi
\ifx \botherref \undefined \def \botherref #1{#1} \fi
\ifx \url \undefined \def \url#1{\textsf{#1}} \fi
\ifx \bchapter \undefined \def \bchapter#1{#1} \fi
\ifx \bbook \undefined \def \bbook#1{#1} \fi
\ifx \bcomment \undefined \def \bcomment#1{#1} \fi
\ifx \oauthor \undefined \def \oauthor#1{#1} \fi
\ifx \citeauthoryear \undefined \def \citeauthoryear#1{#1} \fi
\def \endbibitem {}

\bibitem[\protect\citeauthoryear{Andretta and Jones}{1997}]{Andretta97}
\begin{barticle}
\bauthor{\bsnm{Andretta}, \binits{V.}}, \bauthor{\bsnm{Jones}, \binits{H.P.}}:
\byear{1997},
\bjtitle{Astrophys. J.}
\bvolume{489},
\bfpage{375}.
\end{barticle}
\endbibitem

\bibitem[\protect\citeauthoryear{Benevolenskaya, Kosovichev, and
  Scherrer}{2001}]{Benevol01}
\begin{barticle}
\bauthor{\bsnm{Benevolenskaya}, \binits{E.E.}}, \bauthor{\bsnm{Kosovichev},
  \binits{A.G.}}, \bauthor{\bsnm{Scherrer}, \binits{P.H.}}:
\byear{2001},
\bjtitle{Astrophys. J.}
\bvolume{554},
\bfpage{107}.
\end{barticle}
\endbibitem

\bibitem[\protect\citeauthoryear{Bevington and Robinson}{1992}]{Bevington92}
\begin{bbook}
\bauthor{\bsnm{Bevington}, \binits{P.R.}}, \bauthor{\bsnm{Robinson},
  \binits{D.K.}}:
\byear{1992},
\bbtitle{Data Reduction and Error Analysis for the Physical Sciences},
\bedition{2nd} edn.
\bpublisher{McGraw-Hill},
\blocation{New York},
\bfpage{62}.
\end{bbook}
\endbibitem

\bibitem[\protect\citeauthoryear{Beyer}{1991}]{Beyer91}
\begin{bbook}
\beditor{\bsnm{Beyer}, \binits{W.H.}} (ed.):
\byear{1991},
\bbtitle{{CRC} Standard Mathematical Tables and Formulae},
\bedition{29th} edn.
\bpublisher{CRC Press},
\blocation{Boston},
\bfpage{175}.
\end{bbook}
\endbibitem

\bibitem[\protect\citeauthoryear{Bohlin}{1977}]{Bohlin77}
\begin{bbook}
\bauthor{\bsnm{Bohlin}, \binits{J.D.}}:
\byear{1977},
In: \beditor{\bsnm{Zirker}, \binits{J.B.}} (ed.)
\bbtitle{Coronal Holes and High Speed Wind Streams},
\bpublisher{Colorado Associate University Press},
\blocation{Boulder},
\bfpage{27}.
\end{bbook}
\endbibitem

\bibitem[\protect\citeauthoryear{de~Toma and Arge}{2005}]{deToma05}
\begin{bchapter}
\bauthor{\bparticle{de~}\bsnm{Toma}, \binits{G.}}, \bauthor{\bsnm{Arge},
  \binits{C.N.}}:
\byear{2005},
In: \beditor{\bsnm{Sankarasubramanian}, \binits{K.}}, \beditor{\bsnm{Penn},
  \binits{M.}}, \beditor{\bsnm{Pevtsov}, \binits{A.}} (eds.)
\bbtitle{Large-Scale Structures and their Role in Solar Activity}
\bseriesno{346},
\bpublisher{ASP Conference Series},
\blocation{San Francisco},
\bfpage{251}.
\end{bchapter}
\endbibitem

\bibitem[\protect\citeauthoryear{de~Wit}{2006}]{deWit06}
\begin{barticle}
\bauthor{\bparticle{de~}\bsnm{Wit}, \binits{T.D.}}:
\byear{2006},
\bjtitle{Solar Phys.}
\bvolume{239},
\bfpage{519}.
\end{barticle}
\endbibitem

\bibitem[\protect\citeauthoryear{Harvey and Recely}{2002}]{Harvey02}
\begin{barticle}
\bauthor{\bsnm{Harvey}, \binits{K.L.}}, \bauthor{\bsnm{Recely}, \binits{F.}}:
\byear{2002},
\bjtitle{Solar Phys.}
\bvolume{211},
\bfpage{31}.
\end{barticle}
\endbibitem

\bibitem[\protect\citeauthoryear{Henney and Harvey}{2005}]{Henney05}
\begin{bchapter}
\bauthor{\bsnm{Henney}, \binits{C.J.}}, \bauthor{\bsnm{Harvey}, \binits{J.W.}}:
\byear{2005},
In: \beditor{\bsnm{Sankarasubramanian}, \binits{K.}}, \beditor{\bsnm{Penn},
  \binits{M.}}, \beditor{\bsnm{Pevtsov}, \binits{A.}} (eds.)
\bbtitle{Large-Scale Structures and their Role in Solar Activity}
\bseriesno{346},
\bpublisher{ASP Conference Series},
\blocation{San Francisco},
\bfpage{261}.
\end{bchapter}
\endbibitem

\bibitem[\protect\citeauthoryear{Hoeksema}{2008}]{WSO}
\begin{botherref}
\oauthor{\bsnm{Hoeksema}, \binits{J.T.}}:
2008,
The {Wilcox Solar Observatory}.
\url{http://wso.stanford.edu}.
\end{botherref}
\endbibitem

\bibitem[\protect\citeauthoryear{Howard}{2000}]{Allens}
\begin{bchapter}
\bauthor{\bsnm{Howard}, \binits{R.F.}}:
\byear{2000},
In: \beditor{\bsnm{Cox}, \binits{A.N.}} (ed.)
\bbtitle{Allen's Astrophysical Quantities},
\bedition{4th} edn.,
\bpublisher{Springer},
\blocation{New York},
\bfpage{362}.
\bcomment{Chap. 14.9}.
\end{bchapter}
\endbibitem

\bibitem[\protect\citeauthoryear{Kahler and Hudson}{2002}]{Kahler02}
\begin{barticle}
\bauthor{\bsnm{Kahler}, \binits{S.W.}}, \bauthor{\bsnm{Hudson}, \binits{H.S.}}:
\byear{2002},
\bjtitle{Astrophys. J.}
\bvolume{574},
\bfpage{467}.
\end{barticle}
\endbibitem

\bibitem[\protect\citeauthoryear{Malanushenko and Jones}{2005}]{Malan05}
\begin{barticle}
\bauthor{\bsnm{Malanushenko}, \binits{O.V.}}, \bauthor{\bsnm{Jones},
  \binits{H.P.}}:
\byear{2005},
\bjtitle{Solar Phys.}
\bvolume{226},
\bfpage{3}.
\end{barticle}
\endbibitem

\bibitem[\protect\citeauthoryear{Michielsen and Raedt}{2001}]{Michielsen01}
\begin{barticle}
\bauthor{\bsnm{Michielsen}, \binits{K.}}, \bauthor{\bsnm{Raedt},
  \binits{H.D.}}:
\byear{2001},
\bjtitle{Phys. Rep.}
\bvolume{347},
\bfpage{461}.
\end{barticle}
\endbibitem

\bibitem[\protect\citeauthoryear{Moses \textit{et~al.}}{1997}]{Moses97}
\begin{barticle}
\bauthor{\bsnm{Moses}, \binits{D.}}, \bauthor{\bsnm{Clette}, \binits{F.}},
  \bauthor{\bsnm{Delaboudiniere}, \binits{J.P.}}, \bauthor{\bsnm{Artzner},
  \binits{G.E.}}, \bauthor{\bsnm{Bougnet}, \binits{M.}},
  \bauthor{\bsnm{Brunaud}, \binits{J.}}, \bauthor{\bsnm{Carabetian},
  \binits{C.}}, \bauthor{\bsnm{Gabriel}, \binits{A.H.}},
  \bauthor{\bsnm{Hochedez}, \binits{J.F.}}, \bauthor{\bsnm{Miller},
  \binits{F.}}, \bauthor{\bsnm{Song}, \binits{X.}}, \bauthor{\bsnm{Au},
  \binits{B.}}, \bauthor{\bsnm{Dere}, \binits{K.P.}}, \bauthor{\bsnm{Howard},
  \binits{R.A.}}, \bauthor{\bsnm{Kreplin}, \binits{R.}},
  \bauthor{\bsnm{Michels}, \binits{D.J.}}, \bauthor{\bsnm{Defise},
  \binits{J.M.}}, \bauthor{\bsnm{Jamar}, \binits{C.}}, \bauthor{\bsnm{Rochus},
  \binits{P.}}, \bauthor{\bsnm{Chauvineau}, \binits{J.P.}},
  \bauthor{\bsnm{Marioge}, \binits{J.P.}}, \bauthor{\bsnm{Catura},
  \binits{R.C.}}, \bauthor{\bsnm{Lemen}, \binits{J.R.}}, \bauthor{\bsnm{Shing},
  \binits{L.}}, \bauthor{\bsnm{Stern}, \binits{R.A.}}, \bauthor{\bsnm{Gurman},
  \binits{J.B.}}, \bauthor{\bsnm{Neupert}, \binits{W.M.}},
  \bauthor{\bsnm{Newmark}, \binits{J.}}, \bauthor{\bsnm{Thompson},
  \binits{B.}}, \bauthor{\bsnm{Maucherat}, \binits{A.}},
  \bauthor{\bsnm{Portier-Fozzani}, \binits{F.}}, \bauthor{\bsnm{Berghmans},
  \binits{D.}}, \bauthor{\bsnm{Cugnon}, \binits{P.}}, \bauthor{\bsnm{Dessel},
  \binits{E.L.V.}}, \bauthor{\bsnm{Gabryl}, \binits{J.R.}}:
\byear{1997},
\bjtitle{Solar Phys.}
\bvolume{175},
\bfpage{571}.
\end{barticle}
\endbibitem

\bibitem[\protect\citeauthoryear{{Pesnell} \textit{et~al.}}{2000}]{Pesnell2000}
\begin{barticle}
\bauthor{\bsnm{{Pesnell}}, \binits{W.D.}}, \bauthor{\bsnm{{Goldberg}},
  \binits{R.A.}}, \bauthor{\bsnm{{Jackman}}, \binits{C.H.}},
  \bauthor{\bsnm{{Chenette}}, \binits{D.L.}}, \bauthor{\bsnm{{Gaines}},
  \binits{E.E.}}:
\byear{2000},
\bjtitle{J. Geophys. Res.}
\bvolume{105},
\bfpage{22943}.
doi:\doiurl{10.1029/2000JA000091}.
\end{barticle}
\endbibitem

\bibitem[\protect\citeauthoryear{Press \textit{et~al.}}{1996}]{Numerical_Reci}
\begin{bbook}
\bauthor{\bsnm{Press}, \binits{W.H.}}, \bauthor{\bsnm{Teukolsky},
  \binits{S.A.}}, \bauthor{\bsnm{Vetterling}, \binits{W.T.}},
  \bauthor{\bsnm{Flannery}, \binits{B.P.}}:
\byear{1996},
\bbtitle{Numerical Recipes in Fortan 77: The Art of Scientific Computing},
\bedition{2nd} edn.
\bpublisher{Cambridge University Press},
\blocation{Cambridge, MA},
\bfpage{650}.
\end{bbook}
\endbibitem

\bibitem[\protect\citeauthoryear{Schatten}{2005}]{Schatten05}
\begin{barticle}
\bauthor{\bsnm{Schatten}, \binits{K.H.}}:
\byear{2005},
\bjtitle{Geophys. Res. Lett.}
\bvolume{32},
\bfpage{21106}.
doi:\doiurl{10.1029/2005GL024363}.
\end{barticle}
\endbibitem

\bibitem[\protect\citeauthoryear{Schatten \textit{et~al.}}{1978}]{Schatten78}
\begin{barticle}
\bauthor{\bsnm{Schatten}, \binits{K.H.}}, \bauthor{\bsnm{Scherrer},
  \binits{P.H.}}, \bauthor{\bsnm{Svalgaard}, \binits{L.}},
  \bauthor{\bsnm{Wilcox}, \binits{J.M.}}:
\byear{1978},
\bjtitle{Geophys. Res. Lett.}
\bvolume{5},
\bfpage{411}.
\end{barticle}
\endbibitem

\bibitem[\protect\citeauthoryear{{Sheeley}, {Wang}, and
  {Harvey}}{1989}]{Sheeley89}
\begin{barticle}
\bauthor{\bsnm{{Sheeley}}, \binits{N.R.}}, \bauthor{\bsnm{{Wang}},
  \binits{Y.M.}}, \bauthor{\bsnm{{Harvey}}, \binits{J.W.}}:
\byear{1989},
\bjtitle{Solar Phys.}
\bvolume{119},
\bfpage{323}.
\end{barticle}
\endbibitem

\bibitem[\protect\citeauthoryear{{Svalgaard}, {Cliver}, and
  {Kamide}}{2005}]{Svalgaard01}
\begin{barticle}
\bauthor{\bsnm{{Svalgaard}}, \binits{L.}}, \bauthor{\bsnm{{Cliver}},
  \binits{E.W.}}, \bauthor{\bsnm{{Kamide}}, \binits{Y.}}:
\byear{2005},
\bjtitle{Geophys. Res. Lett.}
\bvolume{32},
\bfpage{1104}.
doi:\doiurl{10.1029/2004GL021664}.
\end{barticle}
\endbibitem

\bibitem[\protect\citeauthoryear{Waldmeier}{1981}]{Waldmeier81}
\begin{barticle}
\bauthor{\bsnm{Waldmeier}, \binits{M.}}:
\byear{1981},
\bjtitle{Solar Phys.}
\bvolume{70},
\bfpage{251}.
\end{barticle}
\endbibitem

\end{thebibliography}

\newpage


\clearpage


\begin{table}
\begin{tabular}{lcccc}     
  \hline                   
Feature & $A$ & $B$ & $C$~ & $\omega$ \\
	      &         &         & 	       & deg day$^{-1}$\\
  \hline
Coronal features & 13.46 & -2.99 &  ---  & 10.82 \\
Surface plasma & 14.11 & -1.70 & -2.35 & 10.77\\
Magnetic field & 14.37 & -2.30 & -1.62 & 11.07 \\
Average &       &    &    &  10.88 \\
  \hline
\end{tabular}
\caption{The coefficients used in Equation~(\ref{Rotation-Rate}) and the resulting rotation rate ($\omega$) for each type of feature to determine the polar rotation rate of PCHs (Howard, 2000).}
\label{T-Rotation}
\end{table}

\end{article}
\end{document}